\documentclass{rmf-d}
\usepackage{nopageno,rmfbib,multicol,times,epsf,amsmath,amssymb,cite}
\usepackage[T1]{fontenc} 
\usepackage[]{caption2}
\usepackage{graphicx}
\usepackage{blindtext}
\usepackage{color}
\usepackage{hyperref}
\usepackage{minted}
\def\rmfcornisa{Research OR Education of Physics \hfill\rmf\ {\bf ??} (*?*) ???--???
\hfill MES? A\~NO?}

\def\rmfcintilla{{\it Rev.\ Mex.\ Fis.\/} {\bf ??} (*?*) (????) ???--???}
\clearpage \rmfcaptionstyle 
\begin{document}
\markboth{ RMF Editorial Team    }{ A \LaTeX template for the RMF, RMF-E, SRMF }

%
%
\title{Black hole mass and distance from accretion disk astrophysical observables 
\vspace{-6pt}}
\author{J. R. Fernández-Moreno$^{a}$, A. González-Juárez$^{b}$, and A. Herrera-Aguilar$^{b}$}
\address{$^{a}$Facultad de Ciencias Físico Matemáticas, Benemérita Universidad Autónoma de Puebla, Apartado Postal 1152, Puebla, Puebla, México.}
%
%
\address{$^{b}$Instituto de Física, Benemérita Universidad Autonoma de Puebla, Apdo. Postal J-48, CP 72570, Puebla, México.}
\author{ }
\address{ }
\author{ }
\address{ }
\maketitle
%
%
\recibido{ }{
\vspace{-12pt}}
\begin{abstract}
\vspace{1em} 
%
%
In this work we derive novel analytical expressions for the mass and distance of a Schwarzschild black hole (BH), as well as for the orbital radius of test particles orbiting it, in terms of astrophysical observables measured throughout the entire orbit of the revolving particle. We use a general relativistic method to describe the frequency shifts of photons emitted in the vicinity of a BH by considering two emitters (or two positions of the same emitter) located symmetrically opposite to each other with respect to the observer's line of sight (LOS) when performing measurements along the orbit. Furthermore, the introduction of the redshift rapidity allows us to write independent expressions for the BH mass and its distance to Earth. We also extend our study to the case when astrophysical systems have a peculiar motion and derive the corresponding closed formulas.
\vspace{1em}
\end{abstract}
\keys{ \bf{\textit{Black hole; mass; distance; emitter orbital radius; analytic formulas; redshift rapidity; megamasers; Active Galactic Nuclei.}} \vspace{8pt}}
\begin{multicols}{2}

\section{Introduction}
BHs emerge as theoretical predictions within the framework of general relativity. Due to the nature of BHs, they must be studied through gravitational interactions with their surroundings, as direct observation is impossible. In the quest for observational evidence confirming the existence of BHs in nature, the detection of gravitational waves provides by the LIGO collaboration robust support for the existence of stellar-mass BHs \cite{r1}. Regarding supermassive BHs, research focuses on the study of galactic nuclei. Type I and II Seyfert galaxies present evidence pointing to their existence, strengthened by the imaging of M87 and Sagittarius A* BH shadows \cite{FirstM87, Sagitario}, from the Event Horizon Telescope (EHT) collaboration. On the other hand, despite the large number of massive dark objects identified as candidates to be supermassive BHs ($10^6-10^{11} M_{\odot}$) \cite{Ko, MDO,  A85}, it remains a challenge to unambiguously determine if a compact object is in fact a BH. Therefore, within this context, the characterization of BHs is of current interest for astrophysical research. 

Studying BHs provides valuable insights into the evolution of astrophysical systems \cite{NSS2000, Kormendy2013}, while their intense gravitational fields offer a unique laboratory for testing gravity models. Some of the BH parameters of primary interest are the mass, the spin, the charge, and its distance to Earth, along with some parameters coming from alternative theories of gravity.

A general relativistic method that allows for the estimation of the aforementioned parameters was developed in \cite{Herrera15} (see as well \cite{review} and references therein). The method consists in deriving analytical expressions for the total frequency shift, taking into account gravitational and kinematic contributions, assuming geodesic motion for both the test particle and the photons they emit. Since the frequency shift is typically a very small quantity, analyses often focus on the extreme cases: the redshift and the blueshift observed near the midline or at positions where velocity gain paths of the test particles are maximized, in order to easy its astronomical measurement.

Furthermore, the method can be adapted to describe several astrophysical systems in a given gravitational field, such as the Reissner–Nordström BH, the Schwarzschild-de Sitter metric, the Kerr and Kerr-de-Sitter backgrounds, among others, being Schwarzschild spacetime the simplest one. For a particularly relevant case, this general relativistic approach considers photon sources in circular orbits around a central compact object, orbiting in the equatorial plane. As a result we obtain analytical expressions for the BH mass and its distance to Earth, as well as for the orbital radius of the photons' emitter, in terms of observable quantities: frequency shifts of photons and photon source positions on the sky. For a Schwarzschild BH, the method allows for the estimation of the mass-to-distance ratio \cite{Nucamendi21} only, since the mass always appears divided by the orbital radius of the emitter in the metric and, hence, in the expression for the frequency shift.

With the purpose of obtaining independent expressions for the mass and distance in terms of astrophysical observables, the redshift rapidity has been introduced in \cite{redshift-rapidity}, defined as the derivative of the frequency shift with respect to the emitter's proper time. 
Another study suggested measuring the angular redshift rate (the derivative of the redshift relative to the position of the photon source on the sky or, equivalently, to the aperture angle of a telescope) \cite{SGS2026}. These models require tracking the path of a photon source and measuring the frequency shift of photons for an arbitrary position along the orbit around the BH. Therefore, the procedure can be readily applied to positions near the midline as well as to those close to the LOS of the orbit of the photons' emitter revolving  the BH. A recent work considered the warping of the BH accretion disk in order to disentangle the mass and distance parameters \cite{AA2025} and presented analytical formulas for them in terms of frequency shifts, sky positions of photon sources, and warping disk parameters. Other works have expressed BH parameters in a similar guise for other spacetime metrics \cite{Herrera15, Becerril2016, Lizardo2018, Nucamendi21b, Banerjee22, Mehrab2023, Martinez2024, Morales2024, Mehrab2025}.

The method is particularly well suited for megamaser systems, where supermassive BHs are thought to power active galactic nuclei (AGN) through matter accretion \cite{BH-engine}. In some cases, astrophysical observables indicate the presence of an accretion disk containing water vapor clouds that emit maser radiation at 22.2 GHz. These clouds move at thousands of kilometers per second around the BH and can be observed using very long baseline interferometry (VLBI). The masers are usually arranged in three groups (the systemic one, near the LOS, and two symmetrically  distributed around the midline with highly redshifted and blueshifted photons) \cite{Lo2005, Herrnstein2005, MCPII, MCPIV, MCPV, MCPVI, MCPVIII, MCPIX}. Since some megamaser systems present equatorial circular orbits and are observed edge-on from Earth, they represent an ideal system to apply the method.

In this work, we provide a generalized version of the aforementioned general relativistic method that incorporates frequency shifts of photons emitted at two arbitrary symmetric positions with respect to the observer's LOS along the orbit of the photon source. In contrast to previous works, where mass and distance derivations depended on midline or LOS positioning, our approach avoids specific approximations. Instead, we provide unified expressions that remain valid across different pairs of orbital maser locations. This broader applicability enhances the robustness of the method and opens the possibility of employing redshift measurements across a wider range of positions along the orbit of the photon source revolving the BH. Even more, beyond recovering the approximate formulas from \cite{redshift-rapidity} as limiting cases near the midline and LOS, we present an analytical expression for the orbital radius of the emitter written in terms of the above mentioned observables.

The outline of this work is as follows. In Section 2, we derive the geodesic four-velocity and four-momentum for massive particles and photons in Schwarzschild spacetime, respectively. In section 3, we derive the photon frequency shift and redshift rapidity for circular equatorial orbits around a BH. In section 4, we make use of the frequency shift and the redshift rapidity measured at two symmetric emitting positions with respect to the LOS along the orbit of the photon source in order to obtain analytical expressions for the mass of the BH, its distance to the Earth as well as the orbital radius of the emitter. In section 5, we consider the peculiar motion of the host galaxy and provide an expression for the frequency shift due this motion using a special relativistic boost along with the corresponding analytic formulas for the BH mass, its distance to Earth and the orbital radius of the photons' emitter. Finally, in Section 6, we present the conclusions of this work.

\section{Particle kinematics in the vicinity of BHs}
In this section, we review the geodesic equations of test particles in the Schwarzschild spacetime, specifically the four-velocity components of massive particles orbiting the BH and the four-momentum components of the photons emitted by it. These expressions are obtained by making use of the Killing vectors and the corresponding conserved quantities (for details see \cite{review} and references therein).

The Schwarzschild metric describes the gravitational field produced by a static spherically symmetric point mass distribution \cite{Balasin}. In Schwarzs\-child coordinates $(t,r,\theta,\varphi)$, the line element (using geometrized units $G=1=c$) reads
\begin{equation}
    ds^2=-\left(1-\frac{2M}{r}\right)dt^2+\frac{dr^2}{1-\frac{2M}{r}}+r^2 d\Omega^2 \,.
\end{equation}
This metric has an event horizon at $r=2M$, where the line element is undefined.

The Schwarzschild metric has four Killing vector fields due to its spherical and temporal symmetries. These vector fields are associated with four conserved quantities: the energy and the components of the angular momentum of a particle with respect to the three spatial axes of symmetry. Thanks to the spherical symmetry of the metric, we can further consider particles motion restricted to the equatorial plane ($\theta=\frac{\pi}{2}$). Therefore, we employ only the temporal (\(\xi =(1,0,0,0)\)) and azimuthal (\(\psi =(0,0,0,1)\)) Killing vector fields. Due to the symmetries along these Killing vector fields, the conserved quantities of a free particle are
\begin{equation} \label{conservatives quantities}
    E = - g_{\nu\mu}\xi^\nu U^\mu \, , \qquad L = g_{\nu\mu}\psi^\nu U^\mu ,
\end{equation}
where $E$ and $L$ are the energy and the azimuthal angular momentum per unit mass, respectively, and $U^\mu$ is the four-velocity of massive particles that will be defined below. Similar expressions hold for the four-mo\-men\-tum of photons and their corresponding conserved quantities $E_\gamma$ and $L_\gamma$.

\subsection{Four-velocity of massive particles}
The four-velocity of a massive particle $U^{\mu}$ is defined with the aid of its proper time $\tau$ as
\begin{equation} \label{4velocidad}
    U^{\mu} = (U^{t}, U^{r}, U^{\theta}, U^{\varphi}),\qquad U^{\mu} = \frac{d x^{\mu}}{d\tau} \,,
\end{equation}
where $x^\mu$ indicates the spacetime position of the particle and the four-velocity satisfies the normalization identity. Substituting the conserved quantities \eqref{conservatives quantities} into the four-velocity normalization condition $U^2=-1$ results in an effective non-relativistic energy conservation law
\begin{eqnarray} \nonumber
    \frac{E^2}{2} = \frac{(U^r)^2}{2} + V_{eff} \,,
\end{eqnarray}
with the effective potential
\begin{eqnarray} \nonumber
    V_{eff} = \frac{1}{2} \left(\frac{L^2}{r^2} + 1\right)\left(1 - \frac{2M}{r}\right) \,.
\end{eqnarray}
In order to consider circular orbits we impose $U^r =0$, hence, $\partial_r V_{eff}=0$ ensuring the existence of a minimum of the effective potential. This condition allows us to obtain expressions for the angular momentum $L$ and the energy E in terms of the BH parameters
\begin{eqnarray} \nonumber
    L = \pm\frac{\sqrt{Mr}}{\sqrt{1-\frac{3M}{r}}} \,, \qquad E = \frac{1 - \frac{2M}{r}}{\sqrt{1-\frac{3M}{r}}} \,,
\end{eqnarray}
where the $\pm$ signs of \(L\) correspond to counterclockwise and clockwise orbital motion, respectively.

If we further require the orbits to be stable, the following condition must also hold $\partial_r^2 V_{eff}>0$, implying that $r > 6M$.

From \eqref{conservatives quantities} we obtain
\begin{eqnarray} \label{4velocity components}
    U^{t} = \frac{1}{\sqrt{1-\frac{3M}{r}}} \,, \qquad    U^{\varphi} = \pm\frac{1}{r} \sqrt{\frac{\frac{M}{r}}{1-\frac{3M}{r}}} .
\end{eqnarray}

\subsection{Four-momentum of photons}
For photons, we work with the four-momentum, defined as
\begin{equation} \label{4wavevector}
    k^{\mu} = (k^{t}, k^{r}, k^{\theta}, k^{\varphi}),\qquad  k^{\mu} = \frac{d x^{\mu}}{d\Lambda} \,,
\end{equation}
where $\Lambda$ is an affine parameter and $x^{\mu}$ is the position four-vector of the photon. In an inertial frame, the temporal  and spatial components of $k^\mu$ match with the energy and linear momentum of the photon, respectively.

The four-momentum satisfies the null condition $k^2=0$. By applying the Killing vector fields as in Eq. \eqref{conservatives quantities} we obtain its components expressed in the language of the conserved energy $E_{\gamma}$ and angular momentum $L_{\gamma}$, and the background metric:
\begin{equation} \label{Kr equation}
    (k^{r})^2 = E_{\gamma}^2 - \frac{L_{\gamma}^2}{r^2}\left(1 - \frac{2M}{r}\right) ,
\end{equation}
\begin{equation} \label{4momentum components}
    k^t = \frac{E_{\gamma}}{1-\frac{2M}{r}} \,, \qquad k^{\varphi} = \frac{L_{\gamma}}{r^2}.
\end{equation}

\section{Relativistic Frequency Shift}
We further review the derivation of frequency shift formulas for arbitrary positions of the photons' emitter (details can be found in \cite{redshift-rapidity}). Photons emitted from a test particle orbiting a compact object undergo a frequency shift owing to gravitational effects (gravitational redshift) in addition to the shift caused by the motion of the photon source (kinematic redshift or blueshift). The frequency of a photon can be expressed in terms of its four-momentum and the four-velocity of the emitter or observer as follows \cite{NSS2000}
\begin{eqnarray} \nonumber
    \omega_p = (-k_{\mu} U^{\mu})\left.\right|_p \,,
\end{eqnarray}
where the subscript $p=e,d$, denotes the emission point $e$ or the detection point $d$.

Since the contraction of the two four-vectors is a scalar, the frequency shift of a photon in a Schwarzschild background is defined as the relativistic invariant
\begin{equation} \label{shift definition}
    1 + z_{Schw} = \frac{\omega_e}{\omega_d} = \frac{(g_{\mu\nu}k^{\mu} U^{\nu})\left.\right|_e}{(g_{\mu\nu}k^{\mu} U^{\nu})\left.\right|_d} \,.
\end{equation}
If we consider a static distant detector and an emitter in circular equatorial motion around a compact object we find 
\begin{equation} \label{shift}
    1 + z_{Schw} = \frac{(g_{tt}k^t U^t + g_{\varphi\varphi}k^{\varphi} U^{\varphi})\left.\right|_e}{(g_{tt}k^t U^t)\left.\right|_d} \,.
\end{equation}
By substituting Eqs. \eqref{4velocity components} and \eqref{Kr equation}-\eqref{4momentum components} into Eq. \eqref{shift} and by choosing the counterclockwise sense of motion, we obtain
\begin{equation} \label{frequency shift}
    1+ z_{Schw} = \frac{1}{\sqrt{1-\frac{3M}{r_e}}}\left(1 - \frac{b_{\gamma}}{r_e} \sqrt{\frac{M}{r_e}}\right) \,,
\end{equation}
where the light deflection parameter is defined as $b_{\gamma} = \frac{L_{\gamma}}{E_{\gamma}}$ and $r_e$ is the orbital radius of photon source. This parameter is derived from the null condition $k^2=0$ and reads
\begin{equation} \label{light parameter}
    b_{\gamma} = \pm r_e\sqrt{\frac{1 - \left(\frac{k^r}{E_{\gamma}}\right)^2}{ 1 - \frac{2M}{r}}} \,,
\end{equation}
where the $\pm$ signs correspond to the particle's position either side of the observer's LOS.

The Schwarzschild frequency shift \eqref{frequency shift} can be decomposed in terms of the gravitational contribution (the constant term at a fixed radius) and the kinematic contribution (the term that describes the rotational motion and contains the light deflection parameter) as \cite{review}
\begin{eqnarray} \label{shift contribution}
    1+z_{Schw}= 1 + z_{grav} + z_{kin} .
\end{eqnarray}
At the positions where the emitter approaches its maximal radial velocity (the velocity projection to the LOS), the total frequency shift is maximized and here $k^r=0$. At these positions the emitter orbital radius can be approximated as $r_e \approx \Theta D$, where $\Theta=\sqrt{(x-x_0)^2+(y-y_0)^2}$ with $(x, y)$ the position on the sky of the emitter and $(x_0, y_0)$ the position of the central compact object, rendering the following relation for the redshift and blueshift of photons
\begin{equation} \label{max redshift}
    R = 1+z_{_{Schw,1}} = \frac{1}{\sqrt{1-\frac{3M}{r_e}}}\left(1 + \sqrt{\frac{\frac{M}{r_e}}{1-\frac{2M}{r_e}}}\right) \,,
\end{equation}
\begin{equation} \label{max blueshift}
    B = 1+z_{_{Schw,2}} = \frac{1}{\sqrt{1-\frac{3M}{r_e}}}\left(1 - \sqrt{\frac{\frac{M}{r_e}}{1-\frac{2M}{r_e}}}\right) \,,
\end{equation}
and yielding the mass-to-distance ratio as a function of observable astrophysical quantities
\begin{equation} \label{MD max shift}
    \frac{M}{D} = \frac{\Theta}{2}\left(\frac{RB - 1}{RB}\right) .
\end{equation}

In order to obtain an expression for the BH mass $M$ decoupled from its distance from Earth $D$, we need to incorporate additional observable quantities, like the aforementioned redshift rapidity \cite{redshift-rapidity}, the accretion disk warping \cite{AA2025} or the angular redshift rate \cite{SGS2026}. Thus, we first need to construct an expression for the frequency shift at an {\it arbitrary position} along the orbit of the photon source.

\begin{figure*}[t]
\centering
\includegraphics[scale=.47]{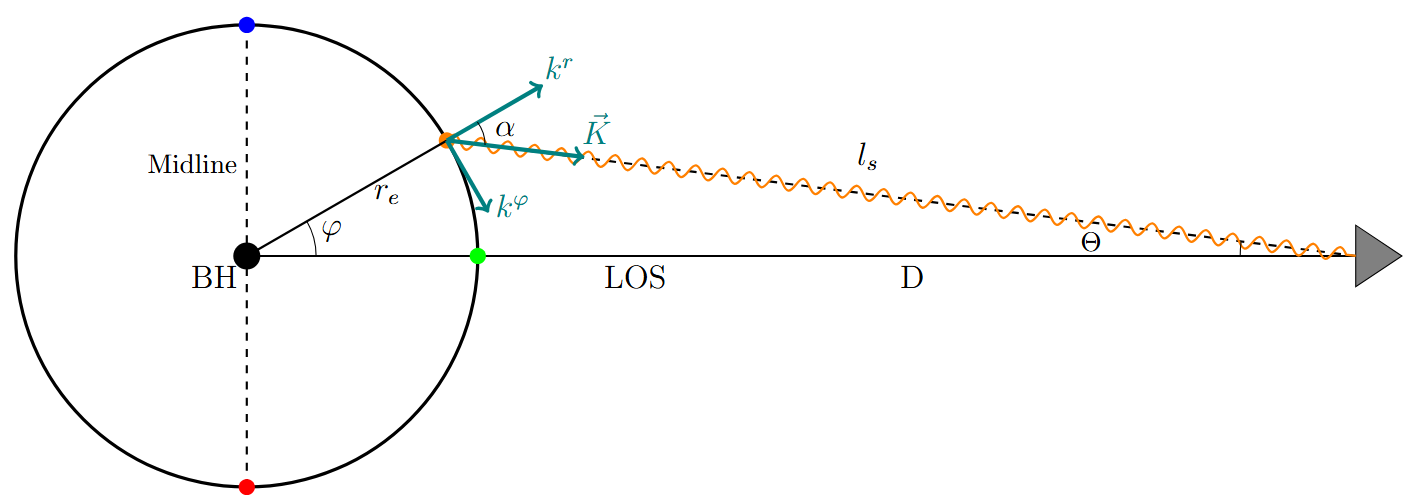}
\caption{An illustration of the relationship between the two-dimensional vector $\vec{K}$ and the components of the four-momentum, $k^r$ and $k^\varphi$, and the origin of the emission angle $\alpha$. We can see that in the Euclidean approximation $\alpha\approx\varphi+\Theta$.}
\label{systemfront}
\end{figure*}

In the frequency shift \eqref{frequency shift}, the contribution due to the motion of the photons' emitter is accounted for in the kinematic term, which contains the light deflection parameter \(b_{\gamma }\). An expression for this parameter can be derived from the restriction $k^2=0$ by considering a two-dimensional vector \(\vec{K}\), which represents the projection of \(k^{\mu }\) onto the equatorial plane\footnote{The emission angle between the radius vector of the emitter and the four-momentum of the photon, $\alpha$, can be approximated as $\alpha\approx\varphi+\Theta$ \cite{redshift-rapidity} in the Euclidean limit.} (see Fig. \ref{systemfront}), as
\begin{equation} \label{deflection parameter}
    b_{\gamma} = - \frac{r_e\sin\alpha}{\sqrt{1-\frac{2M}{r_e}\sin^2\alpha}} .
\end{equation}
Note that the $\pm$ of the relation \eqref{light parameter} is contemplated by the odd property of the function $\sin\alpha$ and the minus sign in \eqref{deflection parameter} has been chosen in order to have physical consistency with the definitions \eqref{max redshift}-\eqref{max blueshift}.
Thus, the frequency shift of photons in the Schwarzschild spacetime with emitters in circular orbits in the equatorial plane around a compact object reads 
\begin{equation} \label{redshift Schw}
    1 + z_{Schw} = \frac{1}{\sqrt{1- 3\tilde{M}}} \left( 1 + \frac{\sqrt{\tilde{M}}\sin \alpha}{\sqrt{1-2\tilde{M}\sin^2\alpha}} \right) \,,
\end{equation}
where we denoted
\begin{eqnarray}
    \tilde{M}= \frac{M}{r_e}\,. 
\end{eqnarray}

\begin{figure*}[t]
\centering
\includegraphics[scale=.8]{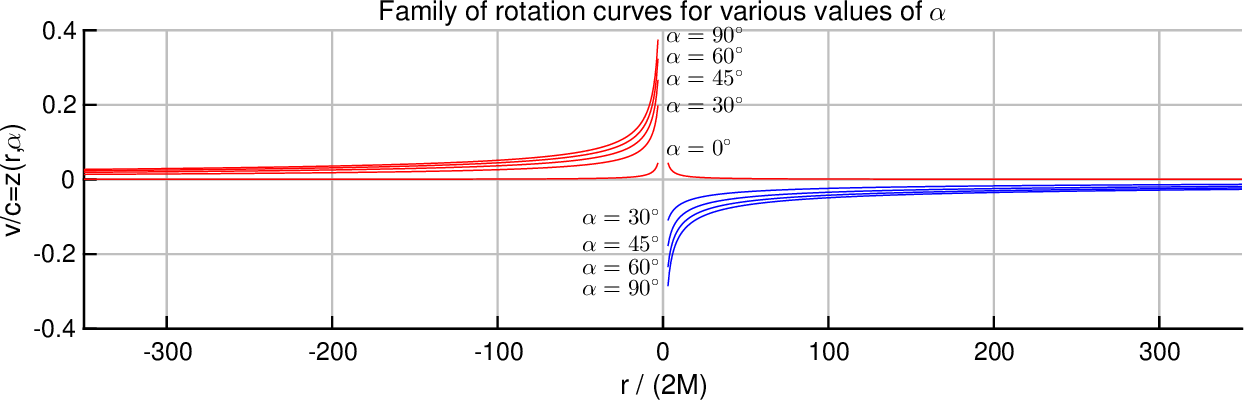}
\caption{Graph of the frequency shifts as functions of the orbital radius $r_e$ (in units of Schwarzschild radii) for different values of the angle $\alpha$. The red curves represent the redshift, while the blue curves correspond to the blueshift. The orbital evolution can be interpreted as a vertical displacement of the curves, from bottom to top, as $\alpha$ varies. In particular, for $\alpha=0$, the redshift and blueshift coincide, corresponding to a purely gravitational redshift. As expected, the magnitude of the frequency shift decreases with increasing distance from the compact object, reflecting the weakening of the gravitational field.}
\label{Zgrap}
\end{figure*}

Rotation curves can be predicted by plotting the frequency shift $z_{Schw}$ with respect to $r_e$ and $\alpha$. Care should be taken not to confuse $\alpha$ with the azimuthal angle $\varphi$ determined by the orbital position of the emitter.
When $\alpha=0$ the source is aligned with the LOS. In that case, the blueshift and redshift curves naturally overlap for the same magnitude of \(r_{e}\), as the geometry of the problem is symmetric. If $r_e$ increases, the curves tend to flatten because the gravitational interaction becomes less intense (see Fig. \ref{Zgrap}).

When the gravitational contribution to the total frequency shift is appreciable, an asymmetry arises between the magnitude of $R$ and $B$: $|R|\neq|B|$. This occurs because the total frequency displacement results from the superposition of gravitational and kinematic terms, in consistency with the expression \eqref{shift contribution}. It thus becomes clear that the kinematic contribution dominates over the gravitational one.

\subsection{Redshift rapidity}
For the astrophysical system under study, there are five unknown quantities: the mass $M$, the orbital radius of the photon source $r_e$, the position of the central compact object on the sky ($x_0,y_0$), and the emission angle $\alpha$, {\it versus}
three astrophysical observables: the location on the sky of the photon emitter ($x,y$) and the frequency shift $z_{Schw}$. Thus, we need to introduce more observable quantities to close the system and be able to extract analytical formulas for the quantities of interest.
Furthermore, from the expression for the frequency shift \eqref{redshift Schw}, it is not possible to decouple $M$ from $r_e$. Recall that at the positions near the midline (where the frequency shift is maximized) $r_e$ can be approximated $r_e\approx\Theta D$ as performed in Eq. \eqref{MD max shift} for systems with a distant observer. In order to enable the decoupling of \(M\) and \(D\) a new observable must be introduced: the redshift rapidity \cite{redshift-rapidity}, defined as the derivative of the frequency shift $z$ with respect to the proper time of the emitter $\tau_e$. Since \(z\) and \(\tau_e\) are relativistic invariants, the derivative \(\frac{dz}{d\tau}\) also does. The redshift rapidity is measured by a distant detector (one that does not experience the BH gravitational field), then the chain rule is implemented to write it in terms of the coordinate time
\begin{equation} \label{dzdt}
    \frac{dz}{d\tau_e} = - \frac{dt}{d\tau_e}\frac{dz}{dt} = \left( U_e^t \right)\frac{dz}{dt} .
\end{equation}
Therefore, by making use of Eqs. \eqref{4velocity components} and 
\eqref{frequency shift} and applying the chain rule to \eqref{dzdt} we find
\begin{eqnarray} \label{dzdt2}
    \frac{dz}{dt} = -\frac{\sqrt{\tilde{M}}}{r_e}\frac{d b_\gamma}{d\alpha}\left(1+\frac{d\Theta}{d\varphi}\right)
    U_e^{\varphi} ,
\end{eqnarray}
where we have also employed the approximation  $\alpha\approx\varphi+\Theta$.

For a distant observer, an expression for the angle $\Theta$ is obtained using the laws of sines and cosines (see Fig. \ref{systemfront})
\begin{eqnarray}
    \frac{l_s}{\sin\varphi} = \frac{r_e}{\sin\Theta},
\end{eqnarray}
\begin{eqnarray}
    l_s^2=D^2+r_e^2-2r_eD\cos\varphi,
\end{eqnarray}
where $l_s$ denotes the distance between the photon source and the observer. Combining these expression, $\cos\Theta$ is written in terms of the parameters of interest ($r_e$, $D$ and $\varphi$). Thus, the angle $\Theta$ is given by
\begin{eqnarray} \label{theta}
    \Theta=\arccos\left( \frac{\tilde{D}-\cos\varphi}{\sqrt{1+\tilde{D}^2 -2\tilde{D}\cos{\varphi}}} \right) \,,
\end{eqnarray}
where
\begin{eqnarray}
    \tilde{D} = \frac{D}{r_e} \,.
\end{eqnarray}
By using Eqs. \eqref{4velocity components}, \eqref{deflection parameter} and \eqref{theta} in \eqref{dzdt2} we obtain
\begin{eqnarray} \label{redshift rapidity}
    &&\frac{dz}{dt} = \\ &&\frac{\tilde{M}\tilde{D}\cos\alpha \left(\tilde{D} - \cos\varphi\right)}{r_e\sqrt{1-3\tilde{M}}\left(1+\tilde{D}^2-2\tilde{D}\cos\varphi\right)\left(1-2\tilde{M}\sin^2\alpha\right)^{\frac{3}{2}}}.\nonumber
\end{eqnarray}

In this expression, $r_e$ appears independently of the ratio $\tilde{M}$. This decoupling allows us to construct independent expressions, first for the emitter radius $r_e$ and subsequently for the BH mass $M$. By expressing $\sin\alpha$ from Eq. \eqref{redshift Schw} and substituting it into \eqref{redshift rapidity}, we obtain a relation involving only the parameters $M$, $r_e$, $x_0$, $y_0$,$\varphi$ and the observables $z$, $\frac{dz}{dt}$, $x$, $y$. Therefore, we consider the frequency shift at a symmetric with respect to the observer's LOS emission position, whose analytical expression involves the same set of parameters, in order to close the system.

\section{Mass, distance and orbital radius at arbitrary emission positions}
To derive an analytical expression for the mass \(M\), we first consider the frequency shift of a photon emitted at two orbital positions. Both positions must be on the same orbit $r_{e_1}=r_{e}=r_{e_2}$, but at opposite angular positions, $\varphi_1=\varphi=-\varphi_2$, with respect to the observer's LOS. Following the notation of Eqs. \eqref{max redshift} and \eqref{max blueshift} for the frequency shifts $R$ and $B$ at the two symmetric locations near the midline, we now define these quantities for two arbitrary but symmetric positions with respect to the LOS along the orbit, that is, using the angular values $\alpha_1=\alpha=-\alpha_2$, for $\alpha>0$, in \eqref{redshift Schw} to write
\begin{align} \label{def R}
    \tilde{R} = 1 + z_{Schw_1} = \frac{1}{\sqrt{1- 3\tilde{M}}} \left( 1 + \frac{\sqrt{\tilde{M}}\sin \alpha}{\sqrt{1-2\tilde{M}\sin^2\alpha}} \right) , \nonumber\\
\end{align}
\begin{align} \label{def B}
    \tilde{B} = 1 + z_{Schw_2} = \frac{1}{\sqrt{1- 3\tilde{M}}} \left( 1 - \frac{\sqrt{\tilde{M}}\sin \alpha}{\sqrt{1-2\tilde{M}\sin^2\alpha}} \right) . \nonumber\\
\end{align}
By performing the sum and subtraction of the frequency shifts we obtain
\begin{equation} \label{sum}
    \tilde{R}+\tilde{B} = \frac{2}{\sqrt{1-3\tilde{M}}} \,,
\end{equation}
\begin{equation} \label{difference}
    \tilde{R}-\tilde{B} = \frac{2}{\sqrt{1-3\tilde{M}}}\left( \frac{\sqrt{\tilde{M}}\sin{\alpha}}{\sqrt{1-2\tilde{M}\sin^2{\alpha}}} \right) \,.
\end{equation}
Due to the even symmetry of the redshift rapidity \eqref{redshift rapidity} with respect to \(\alpha \) and \(\varphi \), the rapidity remains identical for both emission positions
\begin{eqnarray} \nonumber
    \frac{dz}{dt} \Big|_{1} = \frac{dz}{dt} \Big|_{2} = \frac{dz}{dt} .
\end{eqnarray}

Thus, the sum \eqref{sum} can be used to obtain an expression for $\tilde{M}$
\begin{eqnarray} \label{Mtilde}
    \tilde{M}=\frac{\left(\tilde{R} + \tilde{B}\right)^2-4}{3\left(\tilde{R} + \tilde{B}\right)^2} \,,
\end{eqnarray}
and from the subtraction \eqref{difference} we can obtain $\sin\alpha$
\begin{eqnarray} \label{sin}
    \sin\alpha = \sqrt{\frac{3\left(\tilde{R}^2-\tilde{B}^2\right)^2}{\left[(\tilde{R}+\tilde{B})^2-4\right]\left( 3\tilde{R}^2-2\tilde{R}\tilde{B}+3\tilde{B}^2 \right)}} \,.
\end{eqnarray}
Furthermore, by using the law of sines, 
(see Fig. \ref{systemfront}), 
we find
\begin{eqnarray}
    \frac{D}{\sin\alpha} &=& \frac{r_e}{\sin\Theta}, \label{sinus for D}
\end{eqnarray}
\begin{eqnarray}
    \Longrightarrow\tilde{D} &=& \frac{\sin\alpha}{\sin\Theta}, \label{law of sines}
\end{eqnarray}
where $\tilde{D}= D/r_e$. By taking into account the Euclidean approximation, $\alpha\approx\varphi+\Theta$, we express
\begin{eqnarray} \label{cosphi}
    \cos\varphi = \cos\alpha\cos\Theta + \sin\alpha\sin\Theta \,.
\end{eqnarray}
Then by substituting the relations \eqref{sin}, \eqref{law of sines} and \eqref{cosphi} into the redshift rapidity \eqref{redshift rapidity} and solving for $r_e$, we obtain an analytical expression for this quantity that is written in the language of purely astrophysical observables
\begin{eqnarray}\label{radius}
    & &r_e = \frac{\chi^2\zeta\sigma}{ 3\left(\tilde{R}+\tilde{B}\right)^4\left( \frac{dz}{dt} \right)}  \times\nonumber\\
    & & \left[ 2 + \left(\frac{\sin^2\Theta\,\chi^2\zeta^2}{\lambda^2} -1\right) \frac{\lambda\sec^2\Theta}{\lambda +2\tan\Theta\, \sigma} \right]^{-1},\nonumber\\
\end{eqnarray}
where we have defined the following quantities
\begin{eqnarray} \label{coefficients}
    \chi^2 &=& 3\tilde{R}^2-2\tilde{R}\tilde{B}+3\tilde{B}^2\,, \\
    \zeta^2 &=& \left(\tilde{R}+\tilde{B}\right)^2-4 \,,\\ 
    \sigma^2 &=& \left(\tilde{R}+\tilde{B}\right)^2(\tilde{R}\tilde{B}-1) +2\left(\tilde{R}-\tilde{B}\right)^2 \,,\\
    \lambda &=& \sqrt{3}\left(\tilde{R}^2 - \tilde{B}^2\right) \,.
\end{eqnarray}

We further multiply the expressions \eqref{Mtilde} and \eqref{radius} in order to obtain a relation for the mass 
\begin{eqnarray} \label{mass}
    & &M = \frac{\chi^2\zeta^3\sigma}{9\left(\tilde{R}+\tilde{B}\right)^6\left( \frac{dz}{dt} \right)} \nonumber\times\\
    & &\left[ 2 + \left(\frac{\sin^2\Theta\,\chi^2\zeta^2}{\lambda^2} -1\right) \frac{\lambda\sec^2\Theta}{\lambda +2\tan\Theta\, \sigma} \right]^{-1} ,
\end{eqnarray}
and substitute the relations \eqref{sin} and \eqref{radius} into \eqref{sinus for D} to obtain an expression for the BH distance to Earth
\begin{eqnarray}\label{distance}
    & &D = \frac{\left(\tilde{R}-\tilde{B} \right)\chi\sigma}{ \sqrt{3} \left(\tilde{R}+\tilde{B}\right)^3  \sin\Theta\frac{dz}{dt}}\times\nonumber\\
    & & \left[ 2 + \left(\frac{\sin^2\Theta\,\chi^2\zeta^2}{\lambda^2} -1\right) \frac{\lambda\sec^2\Theta}{\lambda +2\tan\Theta\, \sigma} \right] \,,
\end{eqnarray}
relations that again depend on purely astrophysical observable quantities. It is worth to remark that, for a distant observer we have \(\sin\Theta\approx\Theta\) (\(\Theta\ll1\)).

In some astrophysical systems, such as megamasers systems \cite{Humphreys2008}, the redshift rapidity is several orders of magnitude smaller than the frequency shift and vanishes at the locations of maximum frequency shift. In this cases it is convenient to work with the following ratios that are independent of the redshift rapidity
\begin{eqnarray}  \label{MD ratio}
    \frac{M}{D} = \frac{\Theta\,\chi\zeta^3}{3\left(\tilde{R}+\tilde{B}\right)^2\lambda} \,,
\end{eqnarray}
\begin{eqnarray} \label{Dr ratio}
    \frac{D}{r_e} = \frac{1}{\Theta} \frac{\lambda}{\chi\zeta} .
\end{eqnarray}
Notably the mass-to-distance ratio \eqref{MD ratio} reduces to \eqref{MD max shift} at maximized frequency shift positions making use of the relations \eqref{Mtilde} and \eqref{sin} and considering \(\alpha\rightarrow\pm\,\pi/2\).
Besides, the distance-to-radius ratio for a photon source \eqref{Dr ratio} may also prove useful in reconstructing the geometry of the BH accretion disk.

The appearance of the redshift rapidity in the denominator of the expressions for the mass, distance, and orbital radius is inconvenient because it requires greater precision to be measured, as it vanishes at the locations of maximum frequency shift. For this reason, it is convenient to consider emission positions where the redshift rapidity is maximized; however, they coincide with minima of the frequency shift. Moreover, as these locations lie close to the LOS, $\Theta\rightarrow0$ near them, which is particularly troublesome for the distance expression $D$ \eqref{distance}. To overcome this issue, one may use the relations \eqref{MD ratio} and \eqref{Dr ratio} that are independent of the redshift rapidity. 

In addition, we introduce an alternative procedure to determine the BH distance to Earth $D$. This procedure consist of considering two pairs of symmetric emission positions with respect to the observer's LOS: one pair close to the LOS and another one near the midline. By calculating the mass $M$ \eqref{mass} from a pair of photon source locations close to the LOS, then computing the mass-to-distance ratio $M/D$ \eqref{MD ratio} from a pair of symmetric positions near the midline, and further multiplying these relations we obtain an alternative expression for the distance $D$ that reads
\begin{eqnarray} \label{D2}
    & &D_2 = \frac{\left(R_m +B_m\right)^2\left(R_m^2 -B_m^2\right) \chi_L^2 \zeta_L^3\sigma_L}{\Theta_m\frac{dz_m}{dt}\left(R_L +B_L\right)^6 \chi_m\zeta_m^3} \times\nonumber\\
    & & \left[ 2 + \left(\frac{\sin^2\Theta_L\,\chi_L^2\zeta_L^2}{\lambda_L^2} -1\right) \frac{\lambda_L\sec^2\Theta_L}{\lambda_L +2\tan\Theta\, \sigma_L} \right]^{-1}
   ,\nonumber\\
\end{eqnarray}
where the subscript $m$ indicates that the quantities correspond to emission locations near the midline, while the subscript $L$ indicates that they correspond to emission positions near the LOS. The expression \eqref{D2} assumes that each pair of symmetric emission positions shares a common orbital radius, while different pairs may have different radii. We obtain a simpler expression by further assuming that both pairs of emission positions lie along a single orbit
\begin{eqnarray} \label{D3}
   & &D_3 = \frac{\left(R_m - B_m\right) \chi_L^2 \sigma_L}{\Theta_m\frac{dz_m}{dt}\left(R_L +B_L\right)^3 \chi_m } \times\nonumber\\
   & &\left[ 2 + \left(\frac{\sin^2\Theta_L\,\chi_L^2\zeta_L^2}{\lambda_L^2} -1\right) \frac{\lambda_L\sec^2\Theta_L}{\lambda_L +2\tan\Theta_L\, \sigma_L} \right]^{-1}
   .\nonumber\\
\end{eqnarray}

We would like to remark that, unlike the independent expressions for $M$ and $D$ obtained in \cite{redshift-rapidity,SGS2026,AA2025}, 
those found in this work do not require any kind of approximations used for midline and LOS particle positions 
nor introducing new astrophysical observables, like the redshift rapidity and the angular redshift rate, or having a warped accretion disk orbiting the BH.

\section{Peculiar motion}
Several galaxies lodging BH candidates at their core exhibit a peculiar motion, meaning they move due to local gravitational influences from neighboring galaxy clusters. Since this motion is not caused by the expansion of the universe, it is unnecessary to use a metric that accounts for cosmological expansion. Therefore, this peculiar motion can be described by a Lorentz boost \cite{review, Nucamendi21} assuming flat spacetime in the neighborhood of the galaxy hosting the BH.

Consider a emitter system with a constant peculiar velocity with an arbitrary direction relative to the detector system. The total frequency shift is \cite{Tamara2014}
\begin{equation}
    1+z_{total} = \frac{\omega_e}{\omega_d}\frac{{\omega_e}'}{{\omega_d}'} = \left( 1+z_{Schw} \right)\left( 1+z_{boost} \right) \,,
\end{equation}
\begin{equation}
    1+z_{boost} = \frac{{\omega_e}'}{{\omega_d}'} \,,
\end{equation}
where $z_{Schw}$ is the frequency shift for a photon emitted in the vicinity of a compact object in Schwarzschild spacetime, and $z_{boost}$ is the kinematic frequency shift due to a special relativistic boost. The equality holds because $\omega_d$ equals $\omega_e'$ when applying a Lorentz boost. Defining $Z_b=1+z_{boost}$, the corresponding frequency shift reads
\begin{equation}
    Z_{b} = \frac{\omega_e'}{\omega_d'} = \frac{1+\beta\cos\kappa}{\sqrt{1-\beta^2}} ,
\end{equation}
where $\kappa$ es the angle between the direction of motion and the LOS, and $\beta$ is the velocity of the galaxy with respect to the observer in units of $c$.

Since the frequency shift due to a boost is a multiplicative factor in the total frequency shift, the analytical expressions for the orbital radius, the mass and the distance in terms of astrophysical observable quantities become
\begin{eqnarray}\label{boost radius}
    &&r_e = \frac{\chi_t^2\zeta_t\sigma_t}{ 3\left(\tilde{R}_t+\tilde{B}_t\right)^4\left( \frac{dz_t}{dt} \right)} \times \nonumber\\
    & &\left[ 2 + \left(\frac{\sin^2\Theta\,\chi_t^2\zeta_t^2}{\lambda_t^2} -1\right) \frac{\lambda_t\sec^2\Theta}{\lambda_t +2\tan\Theta\, \sigma_t} \right]^{-1} \,,\nonumber\\
\end{eqnarray}
\begin{eqnarray} \label{boost mass}
    &&M = \frac{\chi_t^2\zeta_t^3\sigma_t} {9\left(\tilde{R}_t+\tilde{B}_t\right)^6\left( \frac{dz_t}{dt} \right)} \nonumber\times\\
    & & \left[ 2 + \left(\frac{\sin^2\Theta\,\chi_t^2\zeta_t^2}{\lambda_t^2} -1\right) \frac{\lambda_t\sec^2\Theta}{\lambda_t +2\tan\Theta\, \sigma_t} \right]^{-1} \,,\nonumber\\
\end{eqnarray}
\begin{eqnarray}\label{boost distance}
    &&D = \frac{\left(\tilde{R}_t-\tilde{B}_t \right)\chi_t\sigma_t}{ \sqrt{3} \left(\tilde{R}_t+\tilde{B}_t\right)^3  \sin\Theta\frac{dz_t}{dt}}\times\nonumber\\
    & &\left[ 2 + \left(\frac{\sin^2\Theta\,\chi_t^2\zeta_t^2}{\lambda_t^2             } -1\right) \frac{\lambda_t\sec^2\Theta}{\lambda_t +2\tan\Theta\, \sigma_t} \right] \,,\nonumber\\
\end{eqnarray}
where the quantities $\chi_t$, $\zeta_t$, $\sigma_t$, $\lambda_t$, take the form
\begin{eqnarray}
    \chi_t^2 &=& 3\tilde{R}_t^2-2\tilde{R}_t\tilde{B}_t+3\tilde{B}_t^2 \nonumber,\\
    \zeta_t^2 &=& \left(\tilde{R}_t+\tilde{B}_t\right)^2-4Z_b^2 \nonumber,\\ 
    \sigma_t^2 &=& \left(\tilde{R}_t+\tilde{B}_t\right)^2(\tilde{R}_t\tilde{B}_t-Z_b^2) +2Z_b^2\left(\tilde{R}_t-\tilde{B}_t\right)^2 \nonumber,\\
    \lambda_t &=& \sqrt{3}\left(\tilde{R}_t^2 - \tilde{B}_t^2\right) \nonumber.
\end{eqnarray}

On the other hand, the mass-to-distance and distance-to-radius ratios read
\begin{eqnarray}
    \frac{M}{D} = \frac{\Theta\,\chi_t\zeta_t^3}{3\left(\tilde{R}_t+\tilde{B}_t\right)^2\lambda_t} \,,
\end{eqnarray}
\begin{eqnarray}
    \frac{D}{r_e} = \frac{1}{\Theta} \frac{\lambda_t}{\chi_t\zeta_t} .
\end{eqnarray}

Finally, the alternative expressions for the distance between the Earth and the black hole are written as
\begin{eqnarray} \label{D2 boost}
    & & D_2 = \frac{\left(R_{mt} +B_{mt}\right)^2\left(R_{mt}^2 -B_{mt}^2\right) \chi_{Lt}^2 \zeta_{Lt}^3\sigma_{Lt}}{\Theta_{m}\frac{dz_{mt}}{dt}\left(R_{Lt} +B_{Lt}\right)^6 \chi_{mt}\zeta_{mt}^3} \times\nonumber\\
    & & \left[ 2 + \left(\frac{\sin^2\Theta_L\,\chi_{Lt}^2\zeta_{Lt}^2}{\lambda_{Lt}^2} -1\right) \frac{\lambda_{Lt}\sec^2\Theta_L}{\lambda_{Lt} +2\tan\Theta\, \sigma_{Lt}} \right]^{-1}
   ,\nonumber\\
\end{eqnarray}
\begin{eqnarray} \label{D3 boost}
   & &D_3 = \frac{\left(R_{mt} - B_{mt}\right) \chi_{Lt}^2 \sigma_{Lt}}{\Theta_m\frac{dz_{mt}}{dt}\left(R_{Lt} +B_{Lt}\right)^3 \chi_{mt} } \times\nonumber\\
   & & \left[ 2 + \left(\frac{\sin^2\Theta_L\,\chi_{Lt}^2\zeta_{Lt}^2}{\lambda_{Lt}^2} -1\right) \frac{\lambda_{Lt}\sec^2\Theta_L}{\lambda_{Lt} +2\tan\Theta_L\, \sigma_{Lt}} \right]^{-1}
   .\nonumber\\
\end{eqnarray}

\section{Conclusions}
In this paper we have presented a set of new analytic expressions for the mass of a Schwarzschild BH, its distance from Earth, and the orbital radius of photon sources hosted in the BH accretion disk. These expressions are formulated entirely in terms of astrophysical observable quantities, namely the redshift, blueshift and redshift rapidity of photons emitted from two arbitrary symmetric positions with respect to the observer's LOS, together with their positions on the sky. However, the redshift rapidity is typically several orders of magnitude smaller than the other observables requiring high precision in their measurements. To overcome this limitation, we also derive expressions for the mass-to-distance and distance-to-radius ratios that remain valid at arbitrary orbital locations of a photon source and are completely independent of the redshift rapidity.

A key advantage of this formalism is that it does not rely on measurements restricted to positions of extreme frequency shift as in previous studies \cite{redshift-rapidity}, instead, it proves useful at arbitrary positions along the orbit of a test particle revolving a BH.

The method is particularly well suited for, but not limited to, megamaser systems, where emitters typically form three frequency-shifted groups: two highly redshifted and blueshifted sets of photon sources lying around the midline and a group of systemic masers located near the observer's LOS. In some astrophysical systems, approximate spherical symmetry allows for a fruitful application of this approach.

Finally we would like to note that the method applies to any pair of maser positions, symmetrically located with respect to the observer's LOS, whether they be associated to a single emitter observed at different positions or to distinct emitters sharing the same orbital radius.

\section*{Acknowledgements} 

\noindent Authors are grateful to D. Villaraos,
D.A. Martínez-Valera, U. Nucamendi and M. Momennia for
fruitful discussions, to FORDECYT-PRONACES-CONACYT
for support under grant No. CF-MG-2558591, to VIEP-BUAP
as well as to SNII. A.G.-J. acknowledges financial assistance
from SECIHTI through the postdoctoral grant No. 446473 and J. R. F.-M. through the master's grant No. 2117194.

\end{multicols}

\medline
\begin{multicols}{2}
%
\nocite{*}
\bibliographystyle{rmf-style}
\bibliography{ref}
%
%
%

%
%
\end{multicols}
\end{document}